\documentclass[aps,prb,a4paper,preprint,floatfix,superscriptaddress]{revtex4-1}

\usepackage{graphicx}
\usepackage{amssymb,amsmath}
\usepackage{hyperref}
\usepackage{xspace}
\usepackage{color}
\usepackage{fancyhdr}
\definecolor{delftdark}{rgb}{0, .4, .6}

\begin{document}

\title{Large tunable image-charge effects in single-molecule junctions}

\author{Mickael L. Perrin}\affiliation{Kavli Institute of Nanoscience, Delft University of Technology, Lorentzweg 1, 2628 CJ Delft, The Netherlands}
\author{Christopher J.O. Verzijl}\affiliation{Kavli Institute of Nanoscience, Delft University of Technology, Lorentzweg 1, 2628 CJ Delft, The Netherlands}
\author{Christian A. Martin}\affiliation{Kavli Institute of Nanoscience, Delft University of Technology, Lorentzweg 1, 2628 CJ Delft, The Netherlands}
\author{Ahson J. Shaikh}\affiliation{Department of Chemical Engineering, Delft University of Technology, Julianalaan 136, 2628 BL Delft, The Netherlands}
\author{Rienk Eelkema}\affiliation{Department of Chemical Engineering, Delft University of Technology, Julianalaan 136, 2628 BL Delft, The Netherlands}
\author{Jan H. van Esch}\affiliation{Department of Chemical Engineering, Delft University of Technology, Julianalaan 136, 2628 BL Delft, The Netherlands}
\author{Jan M. van Ruitenbeek}\affiliation{Kamerlingh Onnes Laboratory, Leiden University, Niels Bohrweg 2, 2333 CA Leiden, The Netherlands}
\author{Joseph M. Thijssen}\affiliation{Kavli Institute of Nanoscience, Delft University of Technology, Lorentzweg 1, 2628 CJ Delft, The Netherlands}
\author{Herre S.J. van der Zant}\affiliation{Kavli Institute of Nanoscience, Delft University of Technology, Lorentzweg 1, 2628 CJ Delft, The Netherlands}
\author{Diana Duli\'c}\affiliation{Kavli Institute of Nanoscience, Delft University of Technology, Lorentzweg 1, 2628 CJ Delft, The Netherlands}

\begin{abstract}
The characteristics of molecular electronic devices are critically determined by metal-organic interfaces, which influence the arrangement of the orbital levels that participate in charge transport. Studies on self-assembled monolayers (SAMs) show (molecule-dependent) level shifts as well as transport-gap renormalization, suggesting that polarization effects in the metal substrate play a key role in the level alignment with respect to the metal\rq{s} Fermi energy. Here, we provide direct evidence for an electrode-induced gap renormalization in single-molecule junctions. We study charge transport in single porphyrin-type molecules using electrically gateable break junctions. In this set-up, the position of the occupied and unoccupied levels can be followed in situ and with simultaneous mechanical control. When increasing the electrode separation, we observe a substantial increase in the transport gap with level shifts as high as several hundreds of meV for displacements of a few \AA ngstroms. Analysis of this large and tunable gap renormalization with image-charge calculations based on atomic charges obtained from density functional theory confirms and clarifies the dominant role of image-charge effects in single-molecule junctions.
\end{abstract}

\maketitle

In self-assembled monolayers (SAMs), the influence of the molecule-metal interface on the alignment of the molecular orbital level with respect to the Fermi energy of the substrate 	has been extensively studied with UV and X-ray photo-emission spectroscopy (UPS and XPS)\cite{Ishii1999,Koch2008,Braun2009,Hwang2009}, Kelvin probe measurements\cite{Lange2011,Broker2010}, and scanning tunneling spectroscopy\cite{Otsuki2010}. Such measurements have indicated the formation of an interfacial dipole that is associated with substantial work-function shifts\cite{Ishii1999,Koch2008,Hwang2009,Braun2009,Otsuki2010,Lange2011}, which affect all molecular orbitals in a similar way. Several mechanisms causing this interfacial dipole have been identified. In physisorbed systems the compression of the tail of the electron density outside the metal (``pillow'' or ``push-back'' effect) plays an important role, while for chemisorbed systems charge transfer causes, in addition, a surface dipole to be formed near the metal-molecule interface\cite{Ishii1999,Ishii2000,Heimel2008,Braun2009}. Additionally, straining the molecular junction may shift the orbital levels\cite{Bruot2012}; upon stretching or compression of the molecular junction, the shifts of the occupied and unoccupied level were found to be nearly uniform for the frontier orbitals\cite{Romaner2006}.
Finally, the interaction of the (almost) neutral molecule with its own image-charge distribution at zero bias may also lead to a uniform level shift. This effect is present in both physisorbed and chemisorbed systems. 

In contrast to the previously mentioned effects, UPS experiments probing the ionization and electron addition energies for decreasing layer thicknesses\cite{Amy2005} have shown that the occupied levels move up and the unoccupied ones down in energy, -- this is called  `gap renormalization'. Transport gap renormalization has also been observed in single-molecule devices\cite{Kubatkin2003,Osorio2007} and is commonly explained by the formation of image charges in the metal upon addition or removal of electrons from the molecule\cite{Kaasbjerg2008,Thygesen2009,Barr2012}. This effect occurs repeatedly when a current is passing through it and is particularly apparent in molecules that are weakly coupled to the electrodes.

When varying the electrode separation, the molecular orbital levels are therefore subject to a uniform shift, combined with gap renormalization. Hence, distinguishing the dominant trend in single-molecule junctions requires the combination of an adjustable electrode separation with an electrostatic gate. Although mechanical control over molecular conductance has been reported in various studies\cite{Champagne2005,Martin2010a,Otsuki2010,Parks2010,Meisner2011,Toher2011,Bruot2012,Kim2011}, in only very few reports it has been combined with an electrostatic gate \cite{Champagne2005,Martin2010a}. In particular, systematic studies based on explicit monitoring of the dependence of occupied and unoccupied orbital levels on molecule-electrode distance are lacking. \\

\begin{figure*}
  \begin{center}
  \includegraphics[width=\textwidth]{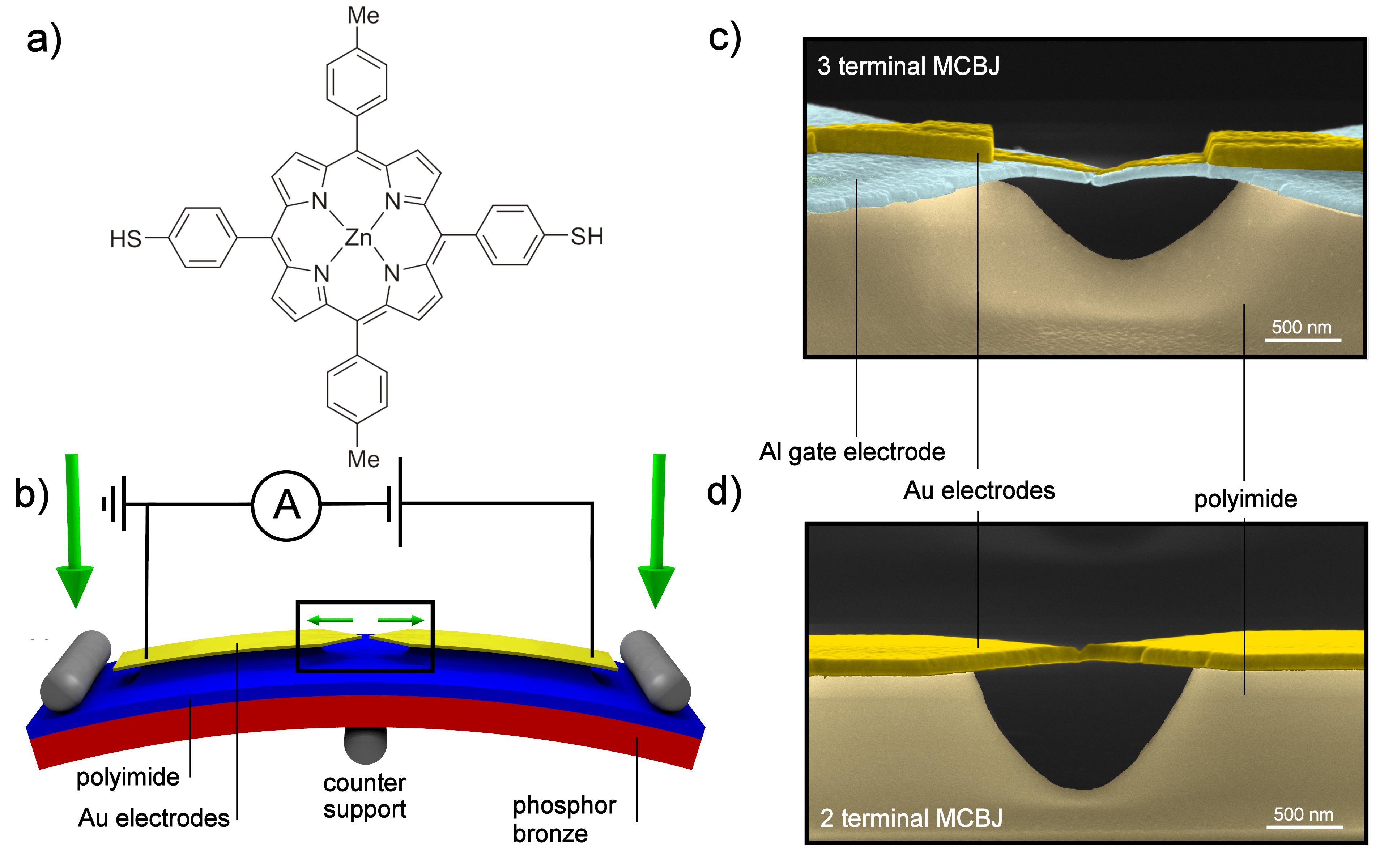}
  \end{center}
 \caption{{\bf Illustration of the experiments.} (a) Structural formula of ZnTPPdT (b) Lay-out of the mechanically controllable break (MCBJ) junction set-up (c) Colorized SEM image of a three-terminal MCBJ device. The gate is made of aluminum and covered with an plasma-enhanced native aluminum oxide layer. The gold electrodes are deposited on top of the gate dielectric (d) Colorized SEM image of a two-terminal MCBJ.}
\label{fig:fig1}
\end{figure*}

{\bf Curent-voltage characteristics}

We have investigated the influence of the metal electrodes on the energy levels in single-molecule junctions using two- and three-terminal mechanically controllable break junctions (MCBJs) in vacuum at 6 K. This architecture (shown in Fig. 1b) allows the distance between the electrodes to be tuned with picometer precision by bending the flexible substrate supporting partially suspended electrodes\cite{Ruitenbeek1996,Kergueris1999,Reichert2002,Dulic2003,Martin2008,Ruben2008,Wu2008}. In three-terminal MCBJ devices an additional gate electrode allows electrostatic tuning of the energy levels of the molecular junction\cite{Martin2010a}. The molecules in this study were thiolated porphyrins, which offer great architectural flexibility and rich optical properties. The thiol-terminated Zn-porphyrin molecules [Zn(5,15-di(p-thiolphenyl)-10,20-di(p-tolyl)porphyrin)], abbreviated as ZnTPPdT and shown in Fig. 1a, were dissolved in dichloromethane (DCM, 0.1 mM) and deposited on the unbroken electrodes using self-assembly from solution. The electrodes were then broken in vacuum at room temperature, cooled down and current-voltage \emph{I-V} characteristics were recorded as a function of electrode spacing. All measurements were performed at 6K. Details concerning these ``systematic I-V series'' and other experimental procedures (synthesis of the molecules, measurement setup, etc.) are provided in the Supplementary Information.

In Fig.~\ref{fig:fig2}a we present typical \emph{I-V} characteristics of a two-terminal MCBJ (sample A) that has been exposed to a solution of ZnTPPdT. We start monitoring the junction-breaking or fusing process at some electrode separation which we call $d_0$. All characteristics show very low current around zero bias, indicating that transport occurs in the weak-coupling (Coulomb-blockade) regime. Steps at higher bias mark the transition to sequential tunneling transport\cite{Cuevas2010}. 
In the differential conductance, \emph{dI/dV}, these steps are visible as peaks (see Fig.~\ref{fig:fig2}b). The peak location identifies the position of the molecular orbital level with respect to the Fermi energy of the electrodes. We will refer to these peaks as resonances from now on. Fig.~\ref{fig:fig2}a,b show that with decreasing inter-electrode distance, the spacing between the resonances is strongly reduced.

\begin{figure*}
  \begin{center}
  \includegraphics[width=\textwidth]{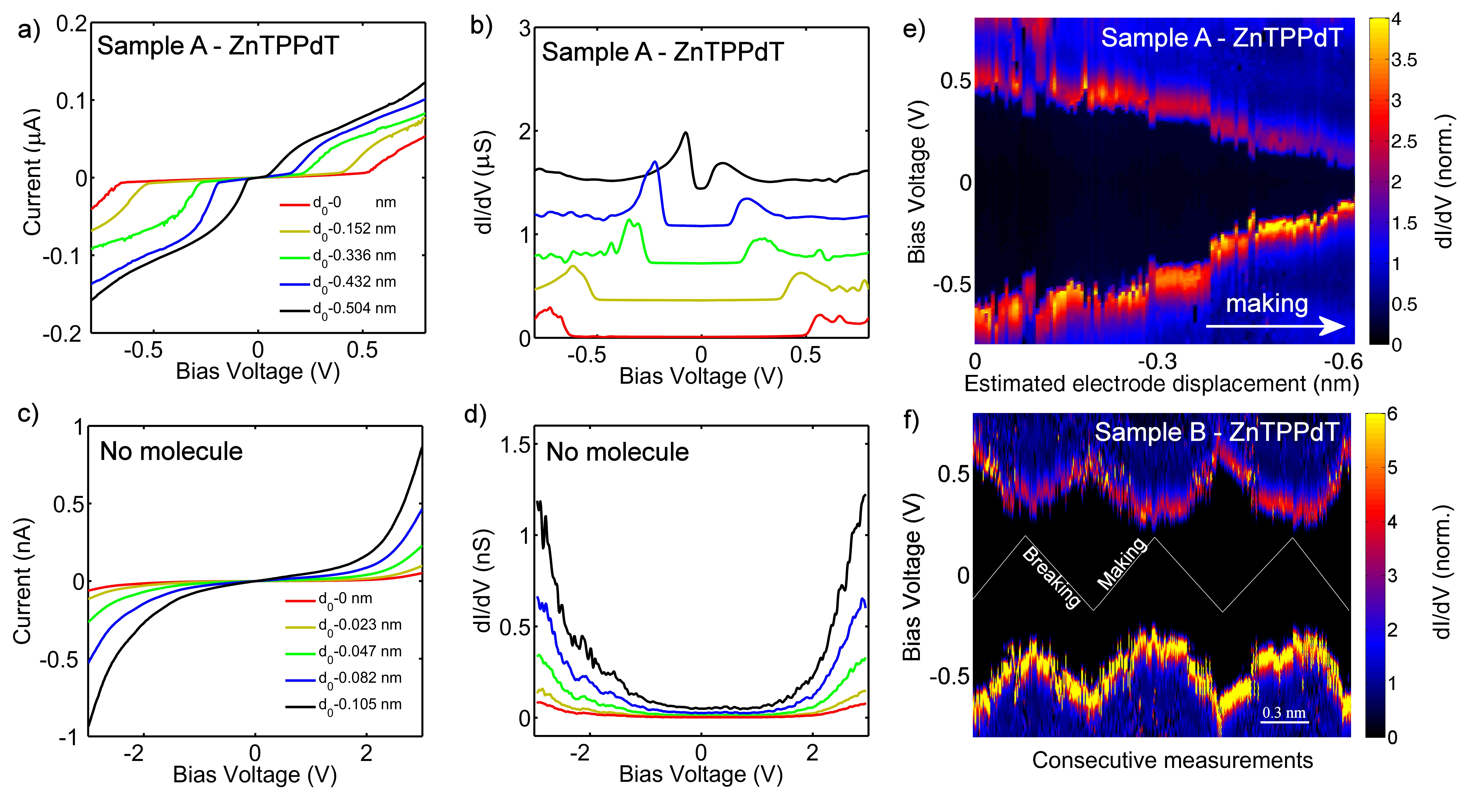}
  \end{center}
  \caption{{\bf Mechanical gating of charge transport in ZnTPPdT junctions.} (a) Current-voltage characteristics and (b) differential conductance for MCBJ devices which have been exposed to a solution of ZnTPPdT. In (c) and (d) the same quantities are plotted for junctions exposed to the pure solvent (DCM). (e,f) Two-dimensional visualization of \emph{dI/dV} for ZnTPPdT as a function of bias voltage and electrode displacement (e) while fusing sample A and (f) for three making/breaking cycles of a different device (sample B). A clear dependence of the Coulomb gap on the electrode spacing is visible. The differential conductance has been normalized and the estimated electrode displacement is relative to $d_0$, the initial electrode separation.}
\label{fig:fig2}
\end{figure*}

We have studied eight different junctions, which all displayed similar mechanically tunable resonances in \emph{dI/dV}. Devices exposed to pure solvent, in contrast, showed featureless characteristics, typical of vacuum tunneling through a single barrier (see Fig.~\ref{fig:fig2}c, 2d). As the inter-electrode distance is reduced, the maximum current in these clean junctions increases smoothly, as a result of the decreasing tunneling barrier width.

To visualize the systematic evolution of the resonance position for hundreds of \emph{dI/dV} curves, we have plotted a two-dimensional map of consecutive I-V measurements in Fig.~\ref{fig:fig2}e, where the gradual shift of the resonances becomes even more apparent. Due to the stability of the electrodes and the fine control over their spacing\cite{Perrin2011,Dulic2009}, the energy levels can be shifted over several hundreds of meV by purely mechanical means. 

In the following, we will refer to these shifts as \emph{mechanical gating} and quantify them in terms of an efficiency factor, the \emph{mechanical gate coupling} (MGC). The MGC is expressed in V/nm and defined as the ratio between the shift of each resonance and the electrode displacement required to achieve this shift. From Fig.~\ref{fig:fig2}e, for example, we find a MGC of about 1 V/nm, with a slight asymmetry for positive and negative bias which may be caused by differences in capacitive coupling to the two electrodes. The reverse process (opening the junction) leads to a widening of the Coulomb gap, as illustrated in Fig.~\ref{fig:fig2}f, where several consecutive opening and closing cycles are shown for a different sample.
The figure clearly shows that the resonances shift consistently and with similar magnitudes, demonstrating the robustness of the effect and the stability of the setup.

While recording systematic I-V series, we occasionally observe a very weak dependence of the resonance positions on the electrode separation, and conversely, occasionally observe MGC's as large as 1.5 V/nm (see Supplementary Information for the statistics of the MGC's). This is probably due to a rearrangement of the molecule inside the junction\cite{Perrin2011,Perrin2011a}.
Alongside gradual changes in the position of the resonances, the plots in Fig.~\ref{fig:fig2}e,f display sudden irreversible jumps in the \emph{dI/dV}'s. These differences and variations could be caused by atomic-scale changes in the geometry of the molecular junction. Evidence of similar rearrangements has also been obtained during room-temperature conductance measurements on porphyrin molecules\cite{Perrin2011a}.
Throughout all the samples, however, the trends remain the same; reducing the electrode distance brings the resonances closer together, whereas increasing the distance moves them further apart.\\

{\bf Gate diagrams}

\begin{figure}
  \begin{center}
  \includegraphics[width=0.87\textwidth]{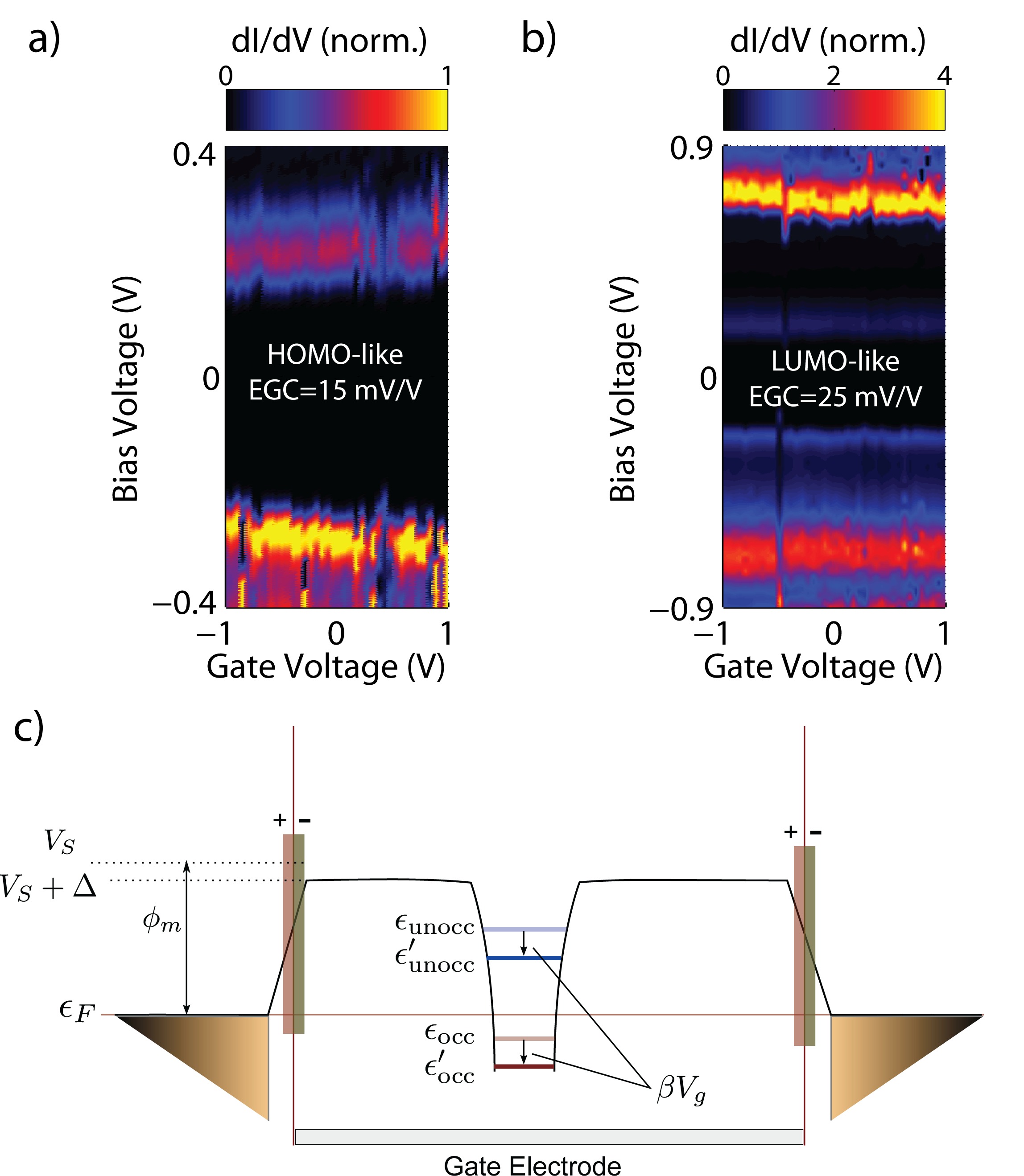}
  \end{center}
 \caption{{\bf Level shifts by electrostatic gating.} (a,b) Gate diagrams recorded on sample C for different junction configurations and during different breaking events. Color-coded $dI/dV$ plotted versus gate- and bias voltage. The slope of the lines allows us to attribute resonances in (a) to an occupied level (HOMO-like, located approx at 0.3eV for zero gate voltage) and those in (b) to an unoccupied level (LUMO-like, located approx at 0.75eV for zero gate voltage). (c) The effect of a rigid shift of the levels under electrostatic gating by a potential $V_g$ applied to a gate electrode below the junction for an occupied and unoccupied level. Here, $\beta$ is the electrostatic gate
coupling, $\phi_m$ the metal work function, $\Delta$ the shift of the potential $V_S$ outside the surface due to the
presence of the molecule, and $\epsilon_F$ the Fermi energy of the metal. $\epsilon_{occ}$, $\epsilon_{unocc}$ and $\epsilon_{occ}'$, $\epsilon_{unocc}'$ are the occupied and unoccupied levels for $V_g=0$ and $V_g\neq 0$, respectively.
 \label{fig:fig3_1}
}
\end{figure}

To obtain additional information about the origin of the shifts of the molecular orbital levels involved in charge transport, we employed electrically gated mechanical break junctions\cite{Martin2010a}. The electrostatic gate in these devices controls the potential on the molecule and lowers/raises all molecular orbital levels for positive/negative gate voltage\cite{Cuevas2010}, as shown in Fig.~\ref{fig:fig3_1}c. Keeping the electrode spacing fixed, we measure the current as a function of both the bias- and gate voltage and plot \emph{dI/dV} as a two-dimensional map; in this paper we will refer to such a plot as a gate diagram.

\begin{figure}
  \begin{center}
  \includegraphics[width=0.8\textwidth]{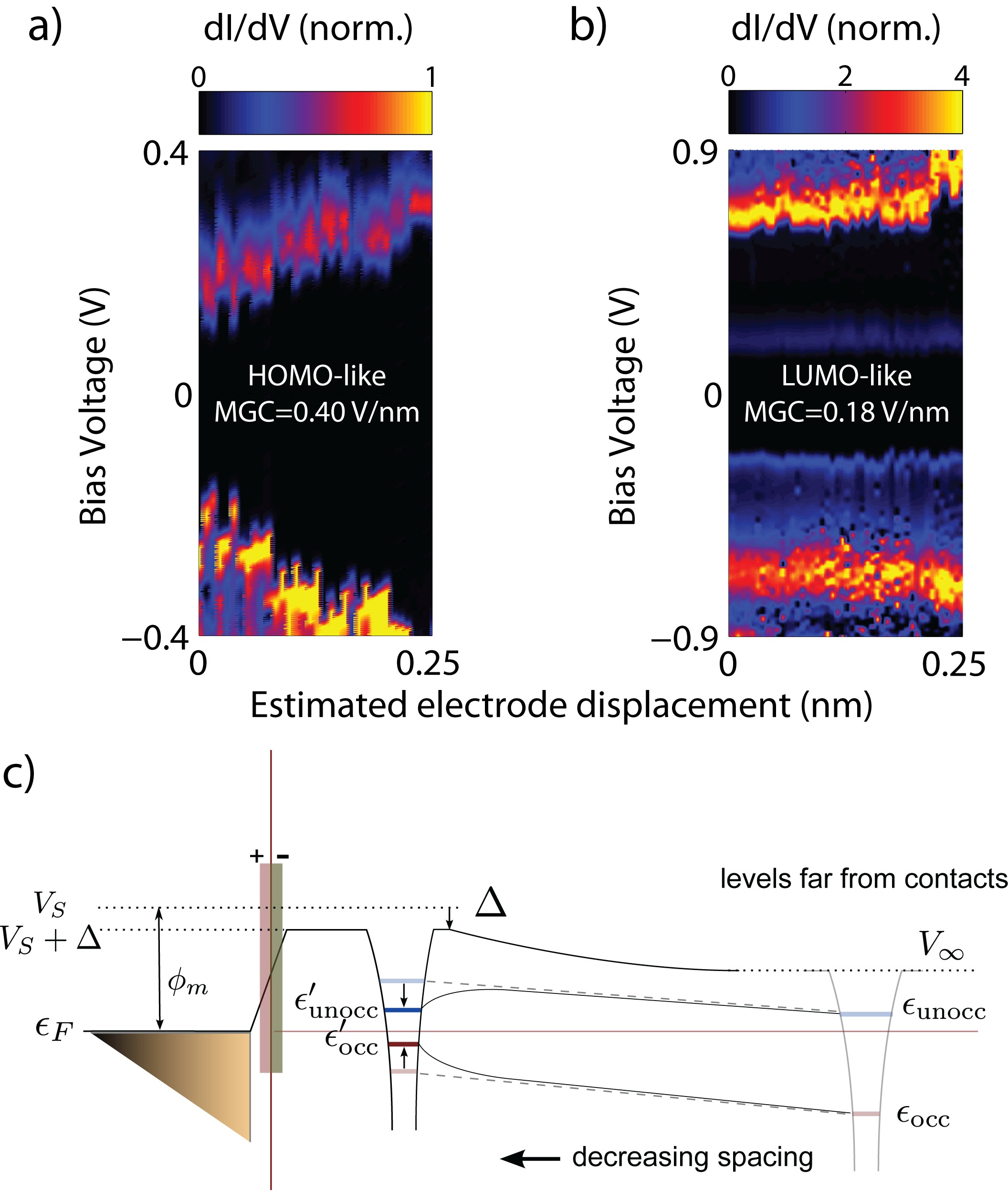}
  \end{center}
  \caption{{\bf Level shifts by mechanical gating.} (a,b) Systematic I-V series of sample C, recorded right after Fig.~\ref{fig:fig3_1} a and b, respectively. (a) HOMO-like (located approx at 0.3eV for zero displacement) and (b) LUMO-like level (located approx at 0.75eV for zero displacement) both move away from the Fermi energy for increasing electrode spacing. (c) The shift of the occupied and unoccupied molecular orbital levels with the distance to the metal. The effects contained in $\Delta$ shift all levels in the same direction, while image-charge effects
are responsible for occupied and unoccupied levels moving closer to the Fermi energy of the metal (gap renormalization). Again, $\phi_m$ represents the metal work function, $\Delta$ the interfacial dipole, $V_\infty$ the potential at infinity, $V_S$ the potential at the surface and $\epsilon_F$ the Fermi energy. $\epsilon_{occ}$, $\epsilon_{unocc}$ and $\epsilon_{occ}'$, $\epsilon_{unocc}'$ are now the occupied and unoccupied levels of the molecule in gas phase and at the interface, respectively.
\label{fig:fig3_2}}
\end{figure}

In such a gate diagram the resonances associated with an occupied level move away from the Fermi level with increasing gate voltage. An unoccupied level, on the other hand, moves closer to the Fermi level and thus displays the opposite trend. This allows us to identify the resonance in Fig.~\ref{fig:fig3_1}a as the HOMO, whereas Fig.~\ref{fig:fig3_1}b shows an unoccupied level (this is not the LUMO of the gas phase molecule, see below). The HOMO level position depends on the gate voltage with an \emph{electrostatic gate coupling} (EGC) of about $15$ mV\, V$^{-1}$; for the unoccupied level, we find an EGC of about $25$ mV\,V$^{-1}$. Fig.~\ref{fig:fig3_2}a--b shows the mechanical gate plots recorded immediately after the measurements shown in Fig.~\ref{fig:fig3_1}b and Fig.~\ref{fig:fig3_1}a, respectively. Both the occupied and unoccupied levels move away from the Fermi level while we increase the distance between the electrodes (MGC $=0.40$ V/nm for the occupied and $0.18$ V/nm for the unoccupied level). 
This implies a widening of the gap and means that the mechanism behind the shifts cannot be a rigid change in the work function only, but must also include a transport-gap renormalization. It is the combination of electrostatic and mechanical gating which leads us to this conclusion, and in the following we will demonstrate using density functional theory (DFT) based calculations that this gap renormalization is caused by the formation of image charges upon charge addition to/removal from the molecule.\\

{\bf DFT calculations}

\begin{figure*}[!h]
 \begin{center}
 \includegraphics[width=\textwidth]{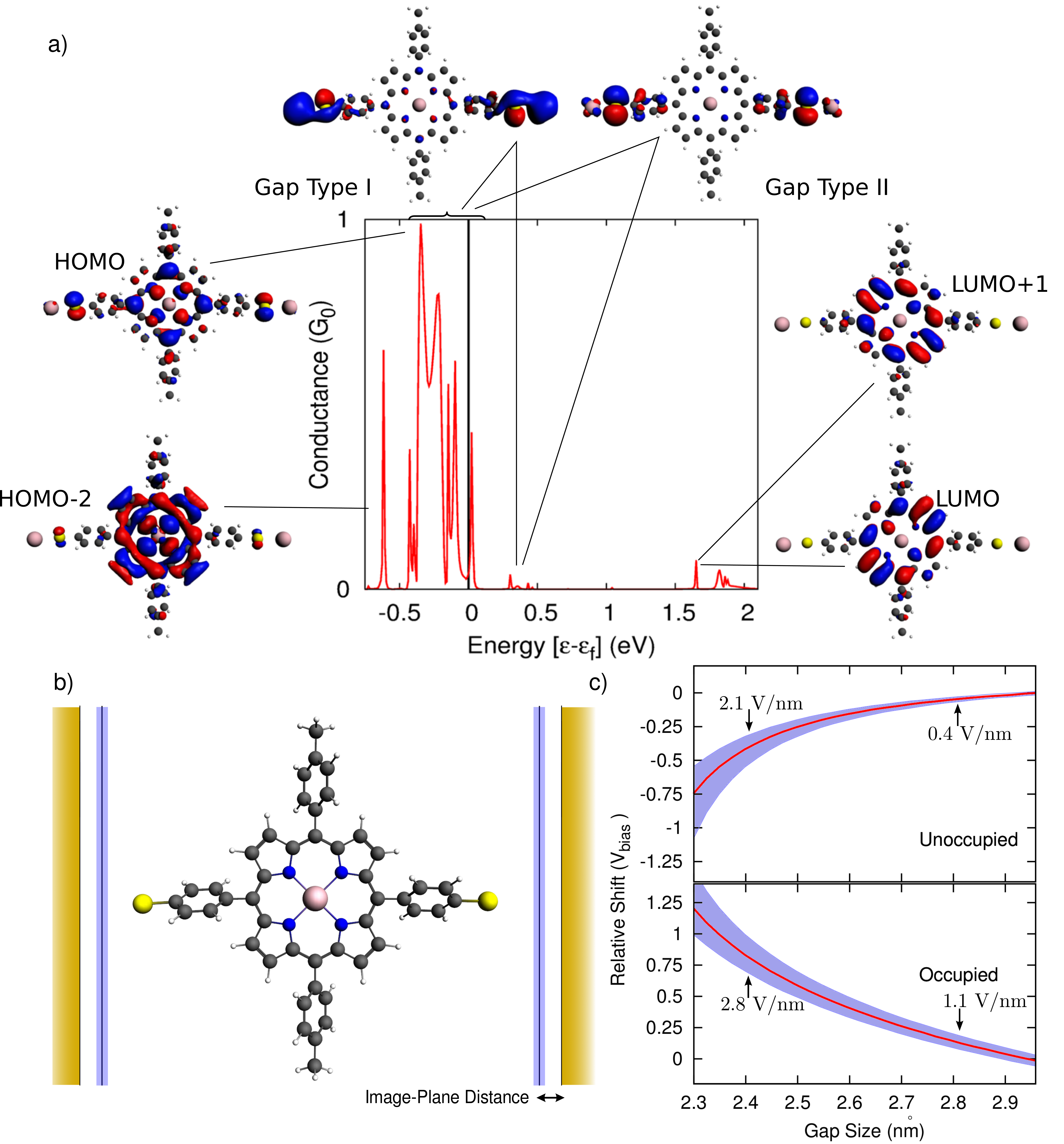}
 \end{center}
\caption{\linespread{1.0}\footnotesize{\bf Transport calculation and image-charge model.} (a) Zero-bias transmission and molecular orbital levels of ZnTPPdT coupled to Au, from DFT and DFT+NEGF calculations, respectively. The Fermi energies are with respect to the Fermi energy of the metal electrodes, marked by the vertical black line. ZnTPPdT is located at 2.59 \AA\xspace from each lead, with hollow-site binding. (b) Image-charge model geometry, with the image plane located 1 \AA\xspace outside the first atomic layer (uncertainty bands derived from a 0.25 \AA\xspace deviation). (c) Shifts predicted by the image-charge model (with uncertainties) showing the occupied- and unoccupied-levels both shifting towards $\epsilon_f$ for decreasing electrode separation with MGC's in the range of 0.4--2.8 V/nm, assuming a symmetrically applied bias. These values may be significantly reduced for realistic electrode geometries.}
\label{fig:fig4}
\end{figure*}

We now turn to the theoretical analysis of the experimentally observed phenomena. Using a quantum chemistry approach\cite{Verzijl2012} we study the electronic structure of the molecules in gas phase and sandwiched between gold atoms in the junctions, as well as their transport properties (see supplementary information for details). In agreement with the literature, our calculations predict a chemisorbed system\cite{Xue2003b,Nara2004,Pontes2011} with ZnTPPdT acting as acceptor, and the hollow site as the most stable configuration. Fig. \ref{fig:fig4} shows the computed zero-bias transmission of the single-molecule junction. We find that the low-bias transport is dominated by the HOMO and HOMO-2 states of the molecule coupled to gold atoms (illustrated in Fig \ref{fig:fig4}a), visible as peaks in the transmission near the Fermi level. 

We also observe that the resonances which correspond to the gas-phase LUMO and LUMO+1 levels are located far above the Fermi level of the leads (although the precise location of these resonances cannot accurately be predicted within DFT). Being more strongly localized at the center of the molecule than the better-hybridizing HOMO-like orbitals, they are expected to have poor conductance properties, and as a consequence are characterized by very narrow peaks in the calculations. A few additional peaks occur slightly above the Fermi energy, and inspection of these states reveals that they have no direct gas-phase counterpart. They are new states, which essentially consist of those parts of the gas-phase HOMO and LUMO that are located on the arms of the molecule and stabilized by the presence of the interface. Forcing an extra electron onto the molecule by applying a positive gate voltage indeed shows that charge is added to these levels, rather than to a LUMO state.

As discussed above, there is a correction $\Delta$ to the background potential which represents a work function shift, as illustrated in Fig.~\ref{fig:fig3_2}c. This shift is usually treated empirically, and is typically negative on Au surfaces. Experiments have reported shifts in the range -0.5 to -1 eV for H$_2$TPP and ZnTPP films\cite{Ishii1999,Ishii2000}, without the presence of the thiols in ZnTPPdT. This corrections is, in principle, distance dependent, and leads to a uniform shift of the occupied and unoccupied levels. This is in contrast with the experimentally observed gap renormalization, indicating that, although this effect may, to some extent, be present, it is not the dominant mechanism responsible for the large level shifts.

Image-charge effects, including their contribution to gap renormalization, can, in principle, be assessed by performing GW calculations\cite{Thygesen2009,Garcia-Lastra2009,Myohanen2012}, which allow for the determination of the ionization potentials and electron addition energies. However, such calculations are infeasible for the large molecules of this study. Instead, we calculate image-charge effects using classical electrostatics based on the atomic charges on the molecule obtained from DFT. In the region where the transport is blocked (corresponding to zero bias and gate) the molecule is approximately, but not exactly, neutral. We call this the `reference state'. The combination of the negatively charged thiols with the positive core of the molecule in the reference state can lead to a contribution of the image charges effect to the uniform shift. This contribution either moves the levels up or down, depending on the exact charge distribution in the junction. To include gap renormalization, one also has to consider the different charged states of the molecule. To access the different charge states in the junction, we added or removed one electron from the molecule by applying a local gate field, in the spirit of a $\Delta-$ SCF method (see supplementary information for details). The image-charge effect corrections are calculated for the different charge states by summing the electrostatic interactions of the atomic charges between two parallel plates with all image charges. The position of the image plane is taken to be $1.0\pm0.25$ \AA\xspace outside the metal surface, as is usually done in the literature\cite{Smith1989,Quek2007,Kaasbjerg2011}. For comparison with experiment, the distance between the electrodes has been varied.

The calculated shifts, illustrated in Fig.~\ref{fig:fig4}c, predict an image-charge contribution to the MGC's in the range of 1.1--2.8 V/nm for an occupied level, and 0.4--2.1 V/nm for an unoccupied level depending on electrode separation. The different molecular orbital levels (shown in Fig.~\ref{fig:fig4}a) thus experience different image charge effects, as observed in the experiments, although the calculated MGC's are larger than the experimental ones. This can be due to the sharp contacts in the MCBJ experiment, which imply a smaller image-charge effect than the large parallel-plate contacts used in the calculation. We modeled the reduction of the image-charge effect with finite contacts, finding it to be roughly a factor of $1.5-2$ (see Supplementary Information), bringing the calculations into better agreement with the experimental shifts. To investigate the sensitivity to the orientation of ZnTPPdT in the junction, we have also rotated the molecule in the calculations and found that the shifts remain essentially the same for angles within 45$^\circ$. 

To assess the contribution to the molecular orbital level shifts originating from structural deformation of the molecule, we performed DFT calculations for increasing gold-gold distance while letting the molecule relax between the contacts. We found that the energy shifts of the occupied and unoccupied level are at most of the order of 50-60~meV, and more importantly, do not lead to transport-gap renormalization, but rather cause an uniform, upward shift. In addition, the HOMO is predicted to move up for increasing electrode-spacing, while the experiments show the opposite trend. 

We conclude that image-charge effects can largely explain the experimentally observed distance dependence of the position of the molecular orbital levels with respect to the Fermi level of the contacts. Our calculations further reveal that the contributions to the image charge effect of the charge distribution in the reference state contributes substantially (roughly half as much as the gap renormalization) to the MGC of the molecular orbital levels.

The time needed for forming image charges is associated with the plasma frequency of the metallic contacts, corresponding to an energy of a few $e$V. This is short enough to be relevant even in co-tunneling processes. In recent years, several attempts have been made to capture the image charge-induced gap renormalization using either single point charges\cite{Neaton2006,Quek2007} or atomic charge distributions\cite{Mowbray2008,Kaasbjerg2008,Kaasbjerg2011}, based on DFT results for gas-phase molecules. In the present system, however, it seems that the states used for electron transport are defined by the presence of the contacts. Therefore, taking the atomic charge distributions for the different charge states inside the junctions is the appropriate starting point for calculating image-charge effects. \\

{\bf Conclusion}

In summary, we have studied the influence of the electrode separation on the molecular orbital levels in porphyrin single-molecule junctions using electrostatically-gated MCBJ devices. Using this method we demonstrate experimentally a combined effect of mechanical and electrostatic gating of the molecular levels. We find that both occupied and unoccupied levels move significantly towards the Fermi level upon reduction of the electrode spacing. We attribute this dominantly to gap renormalization as a result of electron interaction with image charges in the metal leads. Our findings are corroborated by DFT-based calculations. The experiments show surprisingly large level shifts, suggesting that image-charge effects may be responsible for the large spread in conductance values that is often observed in single molecule junctions. These effects should therefore be considered in quantitative comparisons between computations and experiment in single-molecule junctions. At present, calculations for molecular devices result at best in the prediction of trends, or they shed light on the possible transport mechanisms. Improvements in geometric and electrostatic control may bring quantitative agreement between the two closer. We have demonstrated that capturing the image-charge effects is a crucial step in this development. From a different perspective, the observed effects may be exploited to mechanically gate single molecules and thereby tune the alignment of the orbital levels with respect to the Fermi level.

\clearpage
\bibliography{references}

\begin{thebibliography}{10}
\expandafter\ifx\csname url\endcsname\relax
  \def\url#1{\texttt{#1}}\fi
\expandafter\ifx\csname urlprefix\endcsname\relax\def\urlprefix{URL }\fi
\providecommand{\bibinfo}[2]{#2}
\providecommand{\eprint}[2][]{\url{#2}}

\bibitem{Ishii1999}
\bibinfo{author}{Ishii, H.}, \bibinfo{author}{Sugiyama, K.},
  \bibinfo{author}{Ito, E.} \& \bibinfo{author}{Seki, K.}
\newblock \bibinfo{title}{Energy level alignment and interfacial electronic
  structures at organic/metal and organic/organic interfaces}.
\newblock \emph{\bibinfo{journal}{Adv. Mat.}} \textbf{\bibinfo{volume}{11(8)}},
  \bibinfo{pages}{605--625} (\bibinfo{year}{1999}).

\bibitem{Koch2008}
\bibinfo{author}{Koch, N.}
\newblock \bibinfo{title}{Energy levels at interfaces between metals and
  conjugated organic molecules}.
\newblock \emph{\bibinfo{journal}{J. Phys.: Condens. Matter}}
  \textbf{\bibinfo{volume}{20(18)}}, \bibinfo{pages}{184008}
  (\bibinfo{year}{2008}).

\bibitem{Braun2009}
\bibinfo{author}{Braun, S.}, \bibinfo{author}{Salaneck, W.~R.} \&
  \bibinfo{author}{Fahlman, M.}
\newblock \bibinfo{title}{Energy-level alignment at organic/metal and
  organic/organic interfaces}.
\newblock \emph{\bibinfo{journal}{Adv. Mat.}} \textbf{\bibinfo{volume}{21}},
  \bibinfo{pages}{1450--1472} (\bibinfo{year}{2009}).

\bibitem{Hwang2009}
\bibinfo{author}{Hwang, J.}, \bibinfo{author}{Wan, A.} \&
  \bibinfo{author}{Kahn, A.}
\newblock \bibinfo{title}{Energetics of metal-organic interfaces: New
  experiments and assessment of the field}.
\newblock \emph{\bibinfo{journal}{Mater. Sci. Eng. R}}
  \textbf{\bibinfo{volume}{64}}, \bibinfo{pages}{1--31} (\bibinfo{year}{2009}).

\bibitem{Lange2011}
\bibinfo{author}{Lange, I.} \emph{et~al.}
\newblock \bibinfo{title}{Band bending in conjugated polymer layers}.
\newblock \emph{\bibinfo{journal}{Phys. Rev. Lett.}}
  \textbf{\bibinfo{volume}{106}}, \bibinfo{pages}{216402}
  (\bibinfo{year}{2011}).

\bibitem{Broker2010}
\bibinfo{author}{Broker, B.} \emph{et~al.}
\newblock \bibinfo{title}{Density-dependent reorientation and rehybridization
  of chemisorbed conjugated molecules for controlling interface electronic
  structure}.
\newblock \emph{\bibinfo{journal}{Phys. Rev. Lett.}}
  \textbf{\bibinfo{volume}{104}}, \bibinfo{pages}{246805}
  (\bibinfo{year}{2010}).

\bibitem{Otsuki2010}
\bibinfo{author}{Otsuki, J.}
\newblock \bibinfo{title}{Stm studies on porphyrins}.
\newblock \emph{\bibinfo{journal}{Coord. Chem. Rev.}}
  \textbf{\bibinfo{volume}{254}}, \bibinfo{pages}{2311--2341}
  (\bibinfo{year}{2010}).

\bibitem{Ishii2000}
\bibinfo{author}{Ishii, H.} \emph{et~al.}
\newblock \bibinfo{title}{Energy level alignment and band bending at model
  interfaces of organic electroluminescent devices}.
\newblock \emph{\bibinfo{journal}{J. Lumin.}} \textbf{\bibinfo{volume}{61}},
  \bibinfo{pages}{87--89} (\bibinfo{year}{2000}).

\bibitem{Heimel2008}
\bibinfo{author}{Heimel, G.}, \bibinfo{author}{Romaner, L.},
  \bibinfo{author}{Zojer, E.} \& \bibinfo{author}{Bredas, J.-L.}
\newblock \bibinfo{title}{The interface energetics of self-assembled monolayers
  on metals}.
\newblock \emph{\bibinfo{journal}{Acc. Chem. Res.}}
  \textbf{\bibinfo{volume}{41}}, \bibinfo{pages}{721--729}
  (\bibinfo{year}{2008}).

\bibitem{Bruot2012}
\bibinfo{author}{Bruot, C.}, \bibinfo{author}{Hihath, J.} \&
  \bibinfo{author}{Tao, N.}
\newblock \bibinfo{title}{Mechanically controlled molecular orbital alignment
  in single molecule junctions}.
\newblock \emph{\bibinfo{journal}{Nat. Nanotechnol.}}
  \textbf{\bibinfo{volume}{7}}, \bibinfo{pages}{35--40} (\bibinfo{year}{2012}).

\bibitem{Romaner2006}
\bibinfo{author}{Romaner, L.}, \bibinfo{author}{Heimel, G.},
  \bibinfo{author}{Gruber, M.}, \bibinfo{author}{Bredas, J.-L.} \&
  \bibinfo{author}{Zojer, E.}
\newblock \bibinfo{title}{Stretching and breaking of a molecular junction}.
\newblock \emph{\bibinfo{journal}{Small}} \textbf{\bibinfo{volume}{2}},
  \bibinfo{pages}{1468--1475} (\bibinfo{year}{2006}).

\bibitem{Amy2005}
\bibinfo{author}{Amy, F.}, \bibinfo{author}{Chan, C.} \& \bibinfo{author}{Kahn,
  A.}
\newblock \bibinfo{title}{Polarization at the gold/pentacene interface}.
\newblock \emph{\bibinfo{journal}{Org. Electron.}}
  \textbf{\bibinfo{volume}{6}}, \bibinfo{pages}{85--91} (\bibinfo{year}{2005}).

\bibitem{Kubatkin2003}
\bibinfo{author}{Kubatkin, S.} \emph{et~al.}
\newblock \bibinfo{title}{Single-electron transistor of a single organic
  molecule with access to several redox states}.
\newblock \emph{\bibinfo{journal}{Nature}} \textbf{\bibinfo{volume}{425}},
  \bibinfo{pages}{698} (\bibinfo{year}{2003}).

\bibitem{Osorio2007}
\bibinfo{author}{Osorio, E.} \emph{et~al.}
\newblock \bibinfo{title}{Addition energies and vibrational fine structure
  measured in electromigrated single-molecule junctions based on an
  oligophenylenevinylene derivative}.
\newblock \emph{\bibinfo{journal}{Adv. Mat.}} \textbf{\bibinfo{volume}{19(2)}},
  \bibinfo{pages}{281--285} (\bibinfo{year}{2007}).

\bibitem{Kaasbjerg2008}
\bibinfo{author}{Kaasbjerg, K.} \& \bibinfo{author}{Flensberg, K.}
\newblock \bibinfo{title}{Strong polarization-induced reduction of addition
  energies in single-molecule nanojunctions}.
\newblock \emph{\bibinfo{journal}{Nano Lett.}} \textbf{\bibinfo{volume}{8}},
  \bibinfo{pages}{3809--3814} (\bibinfo{year}{2008}).

\bibitem{Thygesen2009}
\bibinfo{author}{Thygesen, K.~S.} \& \bibinfo{author}{Rubio, A.}
\newblock \bibinfo{title}{Renormalization of molecular quasiparticle levels at
  metal-molecule interfaces: Trends across binding regimes}.
\newblock \emph{\bibinfo{journal}{Phys. Rev. Lett.}}
  \textbf{\bibinfo{volume}{102}}, \bibinfo{pages}{046802}
  (\bibinfo{year}{2009}).

\bibitem{Barr2012}
\bibinfo{author}{Barr, J.~D.}, \bibinfo{author}{Stafford, C.~A.} \&
  \bibinfo{author}{Bergfield, J.~P.}
\newblock \bibinfo{title}{Effective field theory of interacting $\pi$
  electrons}.
\newblock \emph{\bibinfo{journal}{Phys. Rev. B}} \textbf{\bibinfo{volume}{86}},
  \bibinfo{pages}{115403} (\bibinfo{year}{2012}).

\bibitem{Champagne2005}
\bibinfo{author}{Champagne, A.}, \bibinfo{author}{Pasupathy, A.} \&
  \bibinfo{author}{Ralph, D.}
\newblock \bibinfo{title}{Mechanically adjustable and electrically gated
  single-molecule transistors}.
\newblock \emph{\bibinfo{journal}{Nano Lett.}} \textbf{\bibinfo{volume}{5}},
  \bibinfo{pages}{305--308} (\bibinfo{year}{2005}).

\bibitem{Martin2010a}
\bibinfo{author}{Martin, C.~A.}, \bibinfo{author}{van Ruitenbeek, J.~M.} \&
  \bibinfo{author}{van~der Zant, H. S.~J.}
\newblock \bibinfo{title}{Sandwich-type gated mechanical break junctions}.
\newblock \emph{\bibinfo{journal}{Nanotechnology}}
  \textbf{\bibinfo{volume}{21}}, \bibinfo{pages}{265201}
  (\bibinfo{year}{2010}).

\bibitem{Parks2010}
\bibinfo{author}{Parks, J.~J.} \emph{et~al.}
\newblock \bibinfo{title}{Mechanical control of spin states in spin-1 molecules
  and the underscreened kondo effect}.
\newblock \emph{\bibinfo{journal}{Science}} \textbf{\bibinfo{volume}{328}},
  \bibinfo{pages}{1370--1373} (\bibinfo{year}{2010}).

\bibitem{Meisner2011}
\bibinfo{author}{Meisner, J.~S.} \emph{et~al.}
\newblock \bibinfo{title}{A single-molecule potentiometer}.
\newblock \emph{\bibinfo{journal}{Nano Lett.}} \textbf{\bibinfo{volume}{11}},
  \bibinfo{pages}{1575--1579} (\bibinfo{year}{2011}).

\bibitem{Toher2011}
\bibinfo{author}{Toher, C.} \emph{et~al.}
\newblock \bibinfo{title}{Electrical transport through a mechanically gated
  molecular wire}.
\newblock \emph{\bibinfo{journal}{Phys. Rev. B}} \textbf{\bibinfo{volume}{83}},
  \bibinfo{pages}{155402} (\bibinfo{year}{2011}).

\bibitem{Kim2011}
\bibinfo{author}{Kim, Y.} \emph{et~al.}
\newblock \bibinfo{title}{Conductance and vibrational states of single-molecule
  junctions controlled by mechanical stretching and material variation}.
\newblock \emph{\bibinfo{journal}{Physical Review Letters}}
  \textbf{\bibinfo{volume}{106}}, \bibinfo{pages}{196804}
  (\bibinfo{year}{2011}).

\bibitem{Ruitenbeek1996}
\bibinfo{author}{van Ruitenbeek, J.~M.} \emph{et~al.}
\newblock \bibinfo{title}{Adjustable nanofabricated atomic size contacts}.
\newblock \emph{\bibinfo{journal}{Rev. Sci. Instrum.}}
  \textbf{\bibinfo{volume}{67}}, \bibinfo{pages}{108--111}
  (\bibinfo{year}{1996}).

\bibitem{Kergueris1999}
\bibinfo{author}{Kergueris, C.} \emph{et~al.}
\newblock \bibinfo{title}{Electron transport through a metal-molecule-metal
  junction}.
\newblock \emph{\bibinfo{journal}{Phys. Rev. B}} \textbf{\bibinfo{volume}{59}},
  \bibinfo{pages}{12505--12513} (\bibinfo{year}{1999}).

\bibitem{Reichert2002}
\bibinfo{author}{Reichert, J.}, \bibinfo{author}{Ochs, R.},
  \bibinfo{author}{Beckmann, H., D.and~Weber}, \bibinfo{author}{Mayor, M.} \&
  \bibinfo{author}{von Lohneysen, H.}
\newblock \bibinfo{title}{Driving current through single organic molecules}.
\newblock \emph{\bibinfo{journal}{Phys. Rev. Lett.}}
  \textbf{\bibinfo{volume}{88}}, \bibinfo{pages}{176804}
  (\bibinfo{year}{2002}).

\bibitem{Dulic2003}
\bibinfo{author}{Duli\'{c}, D.} \emph{et~al.}
\newblock \bibinfo{title}{One-way optoelectronic switching of photochromic
  molecules on gold}.
\newblock \emph{\bibinfo{journal}{Phys. Rev. Lett.}}
  \textbf{\bibinfo{volume}{91}}, \bibinfo{pages}{207402}
  (\bibinfo{year}{2003}).

\bibitem{Martin2008}
\bibinfo{author}{Martin, C.} \emph{et~al.}
\newblock \bibinfo{title}{Fullerene-based anchoring groups for molecular
  electronics}.
\newblock \emph{\bibinfo{journal}{J. Am. Chem. Soc.}}
  \textbf{\bibinfo{volume}{130}}, \bibinfo{pages}{13198--13199}
  (\bibinfo{year}{2008}).

\bibitem{Ruben2008}
\bibinfo{author}{Ruben, M.} \emph{et~al.}
\newblock \bibinfo{title}{Charge transport through a cardan-joint molecule}.
\newblock \emph{\bibinfo{journal}{Small}} \textbf{\bibinfo{volume}{4(12)}},
  \bibinfo{pages}{2229--2235} (\bibinfo{year}{2008}).

\bibitem{Wu2008}
\bibinfo{author}{Wu, S.} \emph{et~al.}
\newblock \bibinfo{title}{Molecular junctions based on aromatic coupling}.
\newblock \emph{\bibinfo{journal}{Nat. Nanotechnol.}}
  \textbf{\bibinfo{volume}{3}}, \bibinfo{pages}{569--574}
  (\bibinfo{year}{2008}).

\bibitem{Cuevas2010}
\bibinfo{author}{Cuevas, J.~C.} \& \bibinfo{author}{Scheer, E.}
\newblock \emph{\bibinfo{title}{Molecular Electronics: An Introduction to
  Theory and Experiment}} (\bibinfo{publisher}{World Scientific},
  \bibinfo{year}{2010}).

\bibitem{Perrin2011}
\bibinfo{author}{Perrin, M.} \emph{et~al.}
\newblock \bibinfo{title}{Charge transport in a zinc-porphyrin single-molecule
  junction}.
\newblock \emph{\bibinfo{journal}{Beilstein J. Nanotechnol.}}
  \textbf{\bibinfo{volume}{2}}, \bibinfo{pages}{714--719}
  (\bibinfo{year}{2011}).

\bibitem{Dulic2009}
\bibinfo{author}{Duli\'{c}, D.} \emph{et~al.}
\newblock \bibinfo{title}{Controlled stability of molecular junctions}.
\newblock \emph{\bibinfo{journal}{Angew. Chem. Int. Ed.}}
  \textbf{\bibinfo{volume}{48(44)}}, \bibinfo{pages}{8273--8276}
  (\bibinfo{year}{2009}).

\bibitem{Perrin2011a}
\bibinfo{author}{Perrin, M.} \emph{et~al.}
\newblock \bibinfo{title}{Influence of the chemical structure on the stability
  and conductance of porphyrin single-molecule junctions}.
\newblock \emph{\bibinfo{journal}{Angew. Chem. Int. Ed.}}
  \textbf{\bibinfo{volume}{50}}, \bibinfo{pages}{11223 --11226}
  (\bibinfo{year}{2011}).

\bibitem{Verzijl2012}
\bibinfo{author}{Verzijl, C.} \& \bibinfo{author}{Thijssen, J.~M.}
\newblock \bibinfo{title}{A dft-based molecular transport implementation in
  adf/band}.
\newblock \emph{\bibinfo{journal}{J. Phys. Chem. C}}  (\bibinfo{year}{2012}).

\bibitem{Xue2003b}
\bibinfo{author}{Xue, Y.} \& \bibinfo{author}{Ratner, M.}
\newblock \bibinfo{title}{Microscopic theory of single-electron tunneling
  through molecular-assembled metallic nanoparticles}.
\newblock \emph{\bibinfo{journal}{Phys. Rev. B}} \textbf{\bibinfo{volume}{68}},
  \bibinfo{pages}{115406} (\bibinfo{year}{2003}).

\bibitem{Nara2004}
\bibinfo{author}{Nara, J.}, \bibinfo{author}{Geng, W.~T.},
  \bibinfo{author}{Kino, H.}, \bibinfo{author}{Kobayashi, N.} \&
  \bibinfo{author}{Ohno, T.}
\newblock \bibinfo{title}{Theoretical investigation on electron transport
  through an organic molecule: Effect of the contact structure}.
\newblock \emph{\bibinfo{journal}{J. Chem. Phys.}}
  \textbf{\bibinfo{volume}{121}}, \bibinfo{pages}{6485--6492}
  (\bibinfo{year}{2004}).

\bibitem{Pontes2011}
\bibinfo{author}{Pontes, R.~B.}, \bibinfo{author}{Rocha, A.~R.},
  \bibinfo{author}{Sanvito, S.}, \bibinfo{author}{Fazzio, A.} \&
  \bibinfo{author}{da~Silva, A. J.~R.}
\newblock \bibinfo{title}{Ab initio calculations of structural evolution and
  conductance of benzene-1,4-dithiol on gold leads}.
\newblock \emph{\bibinfo{journal}{ACS Nano}} \textbf{\bibinfo{volume}{5}},
  \bibinfo{pages}{795--804} (\bibinfo{year}{2011}).

\bibitem{Garcia-Lastra2009}
\bibinfo{author}{Garcia-Lastra, J.~M.}, \bibinfo{author}{Rostgaard, C.},
  \bibinfo{author}{Rubio, A.} \& \bibinfo{author}{Thygesen, K.~S.}
\newblock \bibinfo{title}{Polarization-induced renormalization of molecular
  levels at metallic and semiconducting surfaces}.
\newblock \emph{\bibinfo{journal}{Phys. Rev. B}} \textbf{\bibinfo{volume}{80}},
  \bibinfo{pages}{245427} (\bibinfo{year}{2009}).

\bibitem{Myohanen2012}
\bibinfo{author}{Myohanen, P.}, \bibinfo{author}{Tuovinen, R.},
  \bibinfo{author}{Korhonen, T.}, \bibinfo{author}{Stefanucci, G.} \&
  \bibinfo{author}{van Leeuwen, R.}
\newblock \bibinfo{title}{Image charge dynamics in time-dependent quantum
  transport}.
\newblock \emph{\bibinfo{journal}{Phys. Rev. B}} \textbf{\bibinfo{volume}{85}},
  \bibinfo{pages}{075105} (\bibinfo{year}{2012}).

\bibitem{Smith1989}
\bibinfo{author}{Smith, N.}, \bibinfo{author}{Chen, C.} \&
  \bibinfo{author}{Weinert, M.}
\newblock \bibinfo{title}{Distance of the image plane from metal surfaces}.
\newblock \emph{\bibinfo{journal}{Phys. Rev. B}}
  \textbf{\bibinfo{volume}{40(11)}}, \bibinfo{pages}{7565--7573}
  (\bibinfo{year}{1989}).

\bibitem{Quek2007}
\bibinfo{author}{Quek, S.} \emph{et~al.}
\newblock \bibinfo{title}{Amine--gold linked single-molecule circuits:
  Experiment and theory}.
\newblock \emph{\bibinfo{journal}{Nano Lett.}} \textbf{\bibinfo{volume}{11}},
  \bibinfo{pages}{3477--3482} (\bibinfo{year}{2007}).

\bibitem{Kaasbjerg2011}
\bibinfo{author}{Kaasbjerg, K.} \& \bibinfo{author}{Flensberg, K.}
\newblock \bibinfo{title}{Image charge effects in single-molecule junctions:
  Breaking of symmetries and negative-differential resistance in a benzene
  single-electron transistor}.
\newblock \emph{\bibinfo{journal}{Phys. Rev. B}} \textbf{\bibinfo{volume}{84}},
  \bibinfo{pages}{115457} (\bibinfo{year}{2011}).

\bibitem{Neaton2006}
\bibinfo{author}{Neaton, J.~B.}, \bibinfo{author}{Hybertsen, M.~S.} \&
  \bibinfo{author}{Louie, S.~G.}
\newblock \bibinfo{title}{Renormalization of molecular electronic levels at
  metal-molecule interfaces}.
\newblock \emph{\bibinfo{journal}{Phys. Rev. Lett.}}
  \textbf{\bibinfo{volume}{97}}, \bibinfo{pages}{216405}
  (\bibinfo{year}{2006}).

\bibitem{Mowbray2008}
\bibinfo{author}{Mowbray, D.~J.}, \bibinfo{author}{Jones, G.} \&
  \bibinfo{author}{Thygesen, K.~S.}
\newblock \bibinfo{title}{Influence of functional groups on charge transport in
  molecular junctions}.
\newblock \emph{\bibinfo{journal}{J. Chem. Phys.}}
  \textbf{\bibinfo{volume}{128}}, \bibinfo{pages}{111103}
  (\bibinfo{year}{2008}).

\end{thebibliography}

\begin{itemize}
 \item \textbf{Acknowledgments} This research was carried out with financial support from the Dutch Foundation for Fundamental Research on Matter (FOM),  the VICI (680-47-305) grant from The Netherlands Organisation for Scientific Research (NWO) and the European Union Seventh Framework Programme (FP7/2007-2013) under grant agreement no 270369. The authors would like to thank Ruud van Egmond for expert technical support and Dr. Johannes S. Seldenthuis for fruitful discussions.
\item \textbf{Author Contributions} D.D. and H.v.d.Z. designed the project. C.M., H.v.d.Z. and J.v.R. designed the setup and the devices. M.P. and C.M. fabricated the devices. A.S., R.E. and J.v.E provided the molecules. M.P and D.D. performed the experiments. C.V., M.P. and J.T. performed the calculations. M.P., C.V., D.D., J.T. and H.v.d.Z. wrote the manuscript. All authors contributed to the discussion of the results and the manuscript.
 \item \textbf{Competing Interests} The authors declare that they have no competing financial interests.
 \item \textbf{Correspondence} Correspondence and requests for materials should be addressed to H.S.J. van der Zant~(email: \href{mailto:H.S.J.vanderZant@tudelft.nl}{H.~S.~J.~vanderZant@tudelft.nl}).
\end{itemize}

\end{document}